\documentclass[twocolumn,showplaces,preprintnumbers,amsmath,amssymb,prl,aps]{revtex4-1}

\usepackage{epsfig}
\usepackage{graphicx}
\usepackage{bm}
\usepackage{float} 
\usepackage{amsmath}
\usepackage{gensymb}
\usepackage{epstopdf}

\begin{document}

\title{Anomalies in the core level spectroscopy of a noncentrosymmetric solid, BiPd}

\author{Arindam Pramanik$^1$}
\author{Ram Prakash Pandeya$^1$}
\author{Khadiza Ali$^1$}
\author{Bhanu Joshi$^1$}
\author{Indranil sarkar$^2$}
\author{Paolo Moras$^3$}
\author{Polina M. Sheverdyaeva$^3$}
\author{Asish K. Kundu$^3$}
\author{Carlo Carbone$^3$}
\author{A. Thamizhavel$^1$}
\author{S. Ramakrishnan$^1$}
\author{Kalobaran Maiti$^1$}
\altaffiliation{Corresponding author: kbmaiti@tifr.res.in}

\affiliation{$^1$Department of Condensed Matter Physics and Material Science, Tata Institute of Fundamental Research, Homi Bhabha Road, Colaba, Mumbai - 400005, India \\
$^2$Deutsches Elektronen-Synchrotron DESY, Notkestrasse 85, D-22607 Hamburg, Germany \\
$^3$Istituto di Struttura della Materia, Consiglio Nazionale delle Ricerche, I-34149
Trieste, Italy}


\begin{abstract}
Understanding exotic solids is a difficult task as interactions are often hidden by the symmetry of the system. Here, we study the electronic properties of a noncentrosymmetric solid, BiPd, which is a rare material exhibiting both superconductivity and topological phase of matter. Employing high resolution photoemission spectroscopy with photon energies ranging from hard $x$-ray to extreme ultraviolet regime, we show that hard $x$-ray spectroscopy alone is not enough to reveal surface-bulk differences in the electronic structure. We derived the escape depths close to the extreme surface sensitivity and find that the photon energies used for high resolution measurements such as ARPES fall in the surface sensitive regime. In addition, we discover deviation of the branching ratio of Bi core level features derived from conventional quantum theories of the core hole final states. Such paradigm shift in core level spectroscopy can be attributed to the absence of center of symmetry and spin-orbit interactions.
\end{abstract}

\pacs{31.15.aj, 31.15.em, 61.50.Ah, 71.70.Ej, 79.60.Bm}

\maketitle

\begin{figure}
\vspace{-4ex}
 \begin{center}
 \includegraphics[scale=0.4]{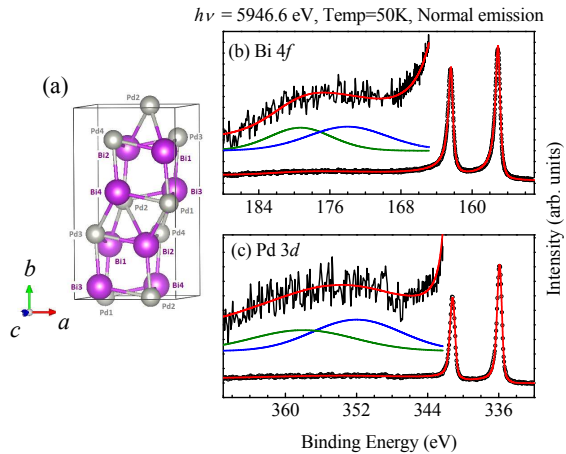}
 \end{center}
 \vspace{-32ex}
\caption{(a) A unit cell of BiPd(monoclinic structure). Inequivalent Bi (Pd) atoms are indexed with 1,2,3 and 4. (b) Bi 4$f$ and (c) Pd 3$d$ HAXPES spectra (open circles). The simulated spectral function is shown by red line superimposed over the experimental data. Insets show the data (noisy line) in an expanded intensity scale exhibiting broad hump, which could be simulated by a set of two peaks separated by the spin-orbit coupling of the corresponding core hole.}
\label{F1-plasmon}
\end{figure}

Recently, non-centrosymmetric superconductors have attracted a lot of attention for holding the possibility of many unusual phenomena ranging from topologically protected surface states to mixing of spin-singlet and spin-triplet components in superconducting pairs, which were otherwise absent in inversion symmetric materials. In this class of materials, BiPd has aroused much interest for being a potential candidate for topological superconductor. BiPd forms in orthorhombic structure at high temperature with the space group $Cmc2_1$, which is called $\beta$-BiPd. Below 483 K, it stabilizes in the monoclinic structure with the space group $P2_1$; it is called $\alpha$-BiPd and it does not have inversion symmetry \cite{Joshi}. In this structure, an unit cell consists of two double layers of Bi and Pd, and both the elements have four nonequivalent sites as shown in Fig.~\ref{F1-plasmon}(a). Single crystals of BiPd have preferential cleaving direction along $b$ axis. Scanning tunneling microscopy showed that cleaved surfaces are Bi terminated, flat and free of reconstruction \cite{Sun}. Small Pd-Pd bond length in this material indicates enhanced Pd 4$d$-4$d$ hybridization, which might play important role in deriving the electronic properties of this material. Various bulk measurements establish an anisotropic superconducting phase below 3.7 K \cite{Joshi-AnisoSC}. Signature of multiple superconducting gap has been observed in point-contact spectroscopy results \cite{Pratap_PRB}. Angle resolved photoemission (ARPES) studies have shown signature of Dirac like surface states in BiPd  \cite{Neupane,Thirupati,Benia}. The spin-resolved photoemission (SPPES) studies have also been carried out to confirm the spin polarization of the surface states \cite{Neupane}.

While exoticity of the material is evident, the surface character of the electronic states have been explained via non-dispersive nature of the corresponding energy bands as a function of photon energy as out-of-plane crystal momentum, $k_z$ is directly related to the photon energy. The non-dispersive nature of the bands indeed relates to the two-dimensional character of the energy bands typical for the surface states. However, similar scenario can also occur due to various other localization effects occurring in the bulk such as electron correlation, disorder, spin/charge order, etc. Thus, such method of the characterization of the surface electronic structure may not be unambiguous. Moreover, in a non-centrosymmetric material, the scenario becomes more complex as the absence of center of symmetry in the bulk crystal structure leads to a potential gradient in the bulk although weaker than that in the surface. Here, we studied the electronic structure of BiPd employing high resolution photoemission spectroscopy with varying surface sensitivity of the technique; this is a direct method of probing the electronic structure at different depths from the sample surface. Experiments with varied photon energies (ultraviolet to hard $x$-ray) reveal an evidence of significant difference between the surface and bulk electronic structures in BiPd. In addition, we discover deviation from conventional quantum behavior of the core level features, which demands physics beyond the existing paradigm of the core level spectroscopy.


High quality single crystals of BiPd were grown using modified Bridgman method. Hard $x$-ray photoemission measurements (HAXPES) were carried out at P09 beamline, PETRA III, Hamburg using a high resolution Phoibos electron analyzer. The setup was optimized for the best energy resolution of 200 meV found at the photon energy of 5946.6 eV. Extreme ultra violet photoemission measurements were performed at the VUV photoemission beamline at Elettra, Trieste using a Gammadata Scienta analyzer, R4000 WAL with an energy resolution set to 15 meV. Sample temperature down to 15 K was controlled using a Helium cryostat. Sample was mounted for easily cleavable $b$ axis after orienting the crystal using Laue diffraction method and was cleaved in situ to expose the clean surface for experiments. Sample surface was found to be clean and reproducibility of the data was verified. The surface sensitivity of the technique was varied in two ways. First, change the electron emission angle keeping the photon energy fixed. As the emission angle is increased, photoelectrons from the bulk have to travel longer path through the sample to reach the sample surface, which diminishes signal intensity from the bulk and consequently, signal from the surface layers gets enhanced. The other method is to vary the photon energy; change in photon energy results in photoemitted electrons with different kinetic energies, and hence different escape depth.


In Fig. \ref{F1-plasmon}(b) and \ref{F1-plasmon}(c), we show the Bi 4$f$ and Pd 4$d$ core level spectra collected at 50 K in normal emission geometry using HAXPES. There are two sharp and distinct asymmetric peaks in each of the spectrum corresponding to the spin-orbit split final states. The asymmetry in the spectral lineshape arises due to the low energy excitations across the Fermi level, which is possible in a system possessing finite density of states at the Fermi level \cite{DoniacSunjic,RuthCore-PRB}. Thus, the observation of asymmetry reflects metallicity of the material. In addition, there is a broad hump in intensity at the higher binding energy side (shown in an enlarged intensity scale in the inset) in both the spectra. The energy separation of the hump from the main peak is quite similar in both the cases indicating a possible origin related to the energy loss in the final state due to the excitations of the collective oscillation modes.

To establish the above assertion, the spectral features are simulated using a set of peaks representing various photoemission signals - the resulting fit (solid line) is superimposed on the experimental spectra (symbols). It is evident that the humps in the spectra could be captured well by two broad peaks separated by the spin-orbit splitting of the main photoemission signals maintaining their intensity ratio; the fit results are shown in the inset. The energy separation of the loss features from the main peak is of the order of 18 eV, which is akin to the bulk plasmon excitation energy found in other systems \cite{DeepSREP}. Similar loss features are also present in other core level spectra. Since, the photoemission process is highly bulk sensitive at 5946.6 eV photon energy, the above features may be considered as bulk property.

\begin{figure}
\vspace{-4ex}
 \begin{center}
 \includegraphics[scale=0.4]{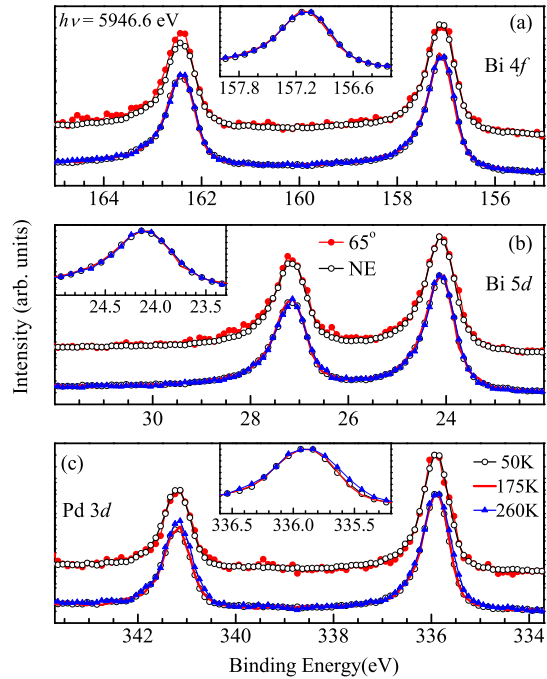}
\end{center}
\vspace{-8ex}
\caption{(a) Bi 4$f$, (b) Bi 5$d$ and (c) Pd 3$d$ core level spectra collected with 5946.6 eV photon energy in normal emission (NE: open circles) and 65$^o$ off-normal emission (solid circles) geometries at a sample temperature of 50 K. In lower panel of each of the plots, normal emission spectra collected at 50 K (open circles), 175 K (line) and 260 K (triangles) are superimposed. Insets show the high angular momentum peak of the temperature dependent data in an expanded energy scale.}
\label{Fig2SurfBulk}
\end{figure}

The temperature dependence of the bulk spectral functions are shown in Fig. \ref{Fig2SurfBulk}; Bi 4$f$, Bi 5$d$ and Pd 3$d$ spectra collected at 50 K, 175 K and 260 K are superimposed in the figure for better comparison. The peak corresponding to the higher $J$ value is shown in the inset in an expanded energy scale. The Bi 4$f$ and Bi 5$d$ spectra are identical at all the temperatures studied. In the case of Pd 3$d$, 50 K and 175 K data are identical but the width is somewhat larger at 260 K. Phonon broadening is expected to manifest in all the core level spectra. However, we do not see broadening induced effect in Bi core level spectra ruling out such possibility. The other possible effect is the proximity to the structural transition; such precursor effect associated to structural transition has been observed in varied materials \cite{precursor}. This can be explained as follows. The width of the core level peak depends on lifetime broadening and screening of the core hole by conduction electrons in the final state. The photoelectron and photohole lifetimes are expected to be less affected by the structural change. Thus, the core level broadening observed here may be attributed to the change in conduction bandwidth as the conduction electrons screening the core hole have larger degree of itineracy in the orthorhombic phase ($\beta$-BiPd) compared to that in the monoclinic phase ($\alpha$-BiPd). The effect is most prominent in the Pd core level as the conduction band is primarily constituted by Pd 4$d$ states; Bi 6$p$ contribution is weak.

In order to investigate the surface-bulk differences, we compare the spectral functions collected at normal emission (NE) and 65$^o$ angled emission geometries in the figure. The escape depth of photoelectrons of kinetic energy close to 6 keV is about 40 \AA. At 65$^o$ off-normal emission, the probing depth will become 20 \AA. In Fig.~\ref{Fig2SurfBulk}, we show the Bi 4$f$, Bi 5$d$ and Pd 3$d$ core level spectra; the data at different emission angles are superimposed. The line shape of all the spectra are very similar indicating its insensitivity to this change of probing depth. This suggests that either the surface and bulk electronic structures are similar or the surface electronic structure, if different, must be limited within the top few layers of the sample that could not be probed by this moderate change in surface sensitivity. It is worth mentioning that no distinct signature of inequivalent Bi and Pd atoms has been observed in the spectra.

All the core level peaks can be simulated with Doniach-$\check{S}$unji\'{c} lineshape using least square error method. Remarkable representation of the asymmetry within the Doniach-$\check{S}$unji\'{c} framework suggests that the asymmetric part originates essentially from the low energy excitations across the Fermi level. Comparable asymmetry in both Bi and Pd core level spectra provide evidence of finite Bi-Pd hybridization \cite{Jong,Folmer}. Interestingly, while the branching ratio of the Pd 3$d$ spin-orbit split peaks is in accordance with the multiplicity of 2:3 of the corresponding final states, Bi core level spectra exhibit significantly higher branching ratio. The lower limit of the ratio is found to be close to 0.71 for Bi 5$d$ and 0.83 for Bi 4$f$ which are significantly higher than the values of 0.67 and 0.75 expected from the multiplicity of the final states - a signature of a paradigm shift of the core level spectroscopy. In the angled emission spectra, this value enhances to 0.74 and 0.86, respectively.

\begin{figure}
\vspace{-4ex}
 \begin{center}
\includegraphics[scale=0.4]{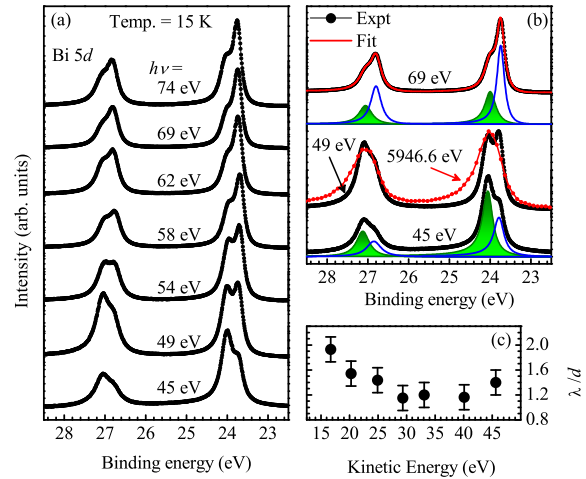}
 \end{center}
 \vspace{-24ex}
 \caption{(a) Bi 5$d$ spectrum collected with different photon energies. (b) Bi 5$d$ spectrum at photon energies 69 eV, 49 eV and 45 eV. We have superimposed HAXPES data over the 49 eV spectrum. The line and shaded area represent the component peaks constituting the experimental spectra - fitting is done following least square error method. (c) The plot of $\lambda/d$ as a function of photoelectron kinetic energy.}
 \label{Fig3-Bi5d}
\end{figure}

Now, we make the technique significantly more surface sensitive using extreme UV radiations and probe the shallow core level, Bi 5$d$; the experimental spectra are shown in Fig.~\ref{Fig3-Bi5d}. Each of the spin-orbit split peak exhibit two distinct features and their relative intensities strongly depend on the photon energy. In Fig. \ref{Fig3-Bi5d}(b), we have superimposed the HAXPES data over the 49 eV spectrum and find that the peak in the hard $x$-ray spectrum matches well with the higher binding energy feature; the linewidth of the HAXPES data is large due to larger resolution and lifetime broadenings of the final states. These observations suggest that the feature at higher binding energy is related to the bulk electronic structure and the other one is the surface feature.

The 45 eV spectrum shows higher intensity of the bulk peak. With the increase in photon energy, the bulk peak becomes gradually weaker and the surface peak stronger; the intensities of the surface and bulk peaks become almost similar at 49 eV and the intensity ratio reverses above 49 eV. We have fit all the spectra using Doniach-$\check{S}$unji\'{c} lineshape representing the features and find a good description in each case - representative fits are shown in Fig. \ref{Fig3-Bi5d}(b). The lineshape of the Bi 5$d$ features are equally broad in bulk and surface. In generally, surface core levels spectra can acquire additional width due to the higher degree of disorder and/or reconstruction occurs at the surface \cite{Newton}. Nearly equal width ($\approx$ 0.3 eV) of the surface and bulk core level peaks suggests that surface-bulk difference in disorder and/or any other surface effects such as reconstructions, impurities are negligible. The intensities of the constituent peaks represent the surface and bulk contribution to the photoemission signal and can be used to calculate the escape depth at different photon energies as discussed below.

The photoemission intensity, $I(\epsilon)$ can be expressed as
$
I(\epsilon) = \int_{0}^{d} I^{s}(\epsilon) e^ {-d/[\lambda(\epsilon)cos\theta]}dx + \int_{d}^{\infty} I^{b}(\epsilon) e^ {-d/[\lambda(\epsilon)cos\theta]}dx.
$
Here, $\lambda(\epsilon)$ is the mean escape depth at kinetic energy $\epsilon$, $d$ is the thickness of the surface layer and $\theta$ is the emission angle with respect to the sample normal. $I^{s}(\epsilon)$ and $I^{b}(\epsilon)$ are the surface and bulk spectral functions, respectively. The first and second terms represent the contributions from surface and bulk electronic structures of the material. The intensity ratio of the surface and bulk contributions can be derived as, $\frac{Surface}{Bulk} = e^{d/\lambda} -1$. Using this relation and the intensity of the component peaks, we estimated the $\lambda /d$ and plotted them in Fig. \ref{Fig3-Bi5d}(c). $\lambda/d$ exhibits a gradual decrease with the increase in photon energy and thereby increase in the kinetic energy of photoelectrons. The minima appears between 35 eV to 40 eV kinetic energy consistent with the findings in various other systems and the universal curve \cite{SeahDench,CSVO,SangeetaSREP}.

\begin{figure}
\vspace{-4ex}
 \begin{center}
\includegraphics[scale=0.4]{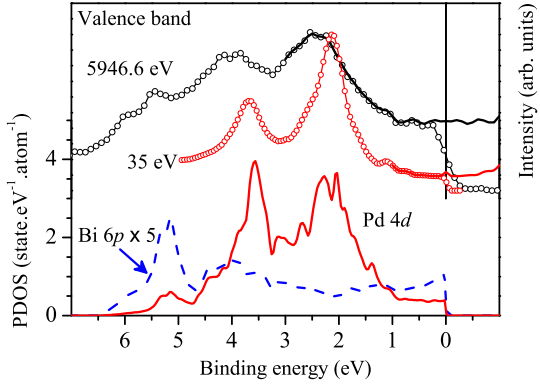}
 \end{center}
 \vspace{-36ex}
 \caption{Valence band spectra of BiPd collected with 5046.6 eV and 35 eV at normal emission. The lines superimposed over the experimental data are the symmetrized data exhibiting flat intensity across the Fermi level. The lines in the lower panel are the calculated partial density of states of Bi 6$p$ (dashed line) and Pd 4$d$ (solid line) constituting the valence band. For clarity, Bi 6$p$ contributions to the formula unit is multiplied by 5.}
 \label{Fig4-VB}
\end{figure}

The valence band spectra collected using hard $x$-ray and 35 eV photon energy is shown in Fig. \ref{Fig4-VB}. The 35 eV spectra corresponds to the electron kinetic energy close to the lowest escape depth and hence essentially provide the surface electronic structure while the hard $x$-ray data represent the bulk electronic structure. For comparison, we show the calculated density of states using density functional theory. The 35 eV data exhibit two sharp peaks around 2.1 eV and 3.7 eV binding energies; these features match remarkably well with the calculated Pd 4$d$ partial density of states (PDOS). We do not see significant contribution from Bi 6$p$ PDOS in the 35 eV spectrum due to dominance of the photoemission cross section of Pd 4$d$ states, $\sigma(Pd4d)$ compared to the cross section of Bi 6$p$ states, $\sigma(Bi6p)$ (the atomic cross section ratio, $\sigma(Bi6p)/\sigma(Pd4d)\approx$ 0.02 \cite{vuo-site}) in addition to the dominant PDOS of Pd 4$d$. Photoemission cross section of Bi 6$p$ becomes appreciable at hard $x$-ray photon energy ($\sigma(Bi6p)/\sigma(Pd4d)\approx$ 0.1 \cite{Yeh}) and the experimental spectra exhibit significant intensity near Fermi level and 5-6 eV binding energy regime.

It is evident from the experimental spectra at two photon energies that the features in the hard $x$-ray data appear at relatively higher binding energies compared to the 35 eV data as also found in the core level data. In order to investigate the spectral function in the vicinity of the Fermi level, we have symmetrized the data with respect to the Fermi level. The results are shown by lines superimposed over the corresponding experimental spectra. In both the cases, the spectral intensity is flat across the Fermi level as expected from the calculated density of states.

\begin{figure}
\vspace{-4ex}
 \begin{center}
\includegraphics[scale=0.4]{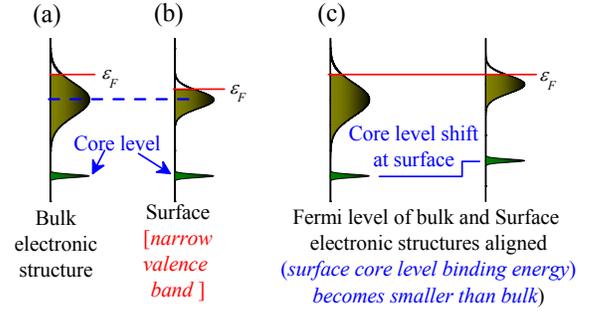}
 \end{center}
 \vspace{-48ex}
 \caption{Schematic exhibiting the valence band and core level; the Fermi level is pinned at the top of the valence band. The surface valence band narrows down due to various surface effects. The alignment of the Fermi level emerges as a lowering of the binding energy of the core level peaks.}
 \label{Fig5-Schem}
\end{figure}

From the above experimental results, one can make three conclusions: (i) the surface and bulk electronic structures of this system are different, (ii) the surface peaks in the core level spectra appear at lower binding energy relative to the bulk features and (iii) the branching ratio of the Bi related core level peaks are much larger than the ratio of the multiplicity of the peaks; the branching ratio of the Pd core level peaks matches well with their multiplicity ratio. Energy shift of the peaks can easily be explained based on surface effects as follows. Due to the translational symmetry breaking along surface normal, the width of the surface energy bands are expected to be narrower than those in the bulk electronic structure as shown in a schematic diagram in Fig. \ref{Fig5-Schem}. Since, the Fermi level is pinned at the top of the Pd 4$d$ band (see calculated results as a reference), a narrowing of the valence band would lead to a shift of the Fermi level to lower energies. The alignment of the Fermi level of the bulk and surface bands leads to a shift of the core level peaks towards lower binding energies.

The angular momentum, $J$ of the core hole created by photoexcitation is $J = L \pm S$; $L$ and $S$ are the orbital and spin quantum number, respectively. The intensity of the peaks corresponding to each of the core hole state will depend on it's multiplicity, (2$J$+1) and hence, the branching ratio is $[2(L-S)+1]:[2(L+S)+1]$. This is observed to hold good in various core level spectroscopy. The deviation from such behavior suggests a paradigm shift from atomic descriptions as the behavior of electrons in solid is expected to be different. The orbital angular momentum gets significantly modified due to crystal momentum in solid. In relativistic description, the spin-orbit term for an electron in an external scaler potential, $V(r)$ can be expressed as $-{{e\hbar}\over{(2mc)^2}}\sigma.(E(r)\times p)$. Here, $E(r)$ is the electric field (= -$\nabla V(r)$) and $p$ is electron momentum. In a crystal without inversion symmetry, the potential gradient and hence, the electric field will be finite, which will lead to appreciable angular momentum. This description helped to capture the electronic structure of non-centrosymmetric semiconductors such as GaAs, InSb, etc. and is known as Dresselhaus effect \cite{Dresselhaus}. Such effect gets significantly enhanced in heavier elements \cite{SOC}.

While core levels believed to follow atomic description, the translational symmetry in the solid will affect the dispersion of the core electronic states although not documented in the literature (to our knowledge). The absence of the center of symmetry in BiPd will lead to an enhancement of the orbital angular momentum and hence the effective $J$ in BiPd will be larger than the atomic value. This effect will be significantly strong for Bi than Pd (atomic mass of Bi is close to double of the atomic mass of Pd). If we take the branching ratio calculated for normal emission spectra, the $L_{eff}$ for Bi 5$d$ is found to be about 2.4 and for 4$f$, it is 4.9; larger increase for 4$f$ electrons indicates their higher sensitivity to the heavier mass of Bi as 4$f$ electrons possess more local character than the 5$d$ electrons. Such scenario is consistent with the observation of enhancement of branching ratio at the surface due to further enhancement of potential gradient at the surface as found in E. I. Rashba's description \cite{Rashba}.


In summary, we studied the electronic structure of a non-centrosymmetric superconductor, BiPd using high resolution photoemission spectroscopy. Experiments were carried out on high quality single crystals using hard $x$-ray and ultraviolet radiations at different experiment topologies. We show that while HAXPES captured the bulk electronic structure well, the distinct differences between the surface and bulk electronic structure could be revealed using UV energies. The surface peaks appear at lower binding energy than the bulk peaks that can be attributed to the narrowing of the surface electronic structure. We estimated the escape depth of photoelectrons as a function of electron kinetic energy exhibiting highest surface sensitivity near 40 eV electron kinetic energy. Most interestingly, we discovered deviation of the branching ratio of Bi core level peaks from the ratio of the final states multiplicity calculated using atomic description, while the Pd core levels exhibit atomic behavior. To our knowledge, for the first time, these results provide an evidence of paradigm shift of the core level spectroscopy requiring to go beyond atomic description.

Financial support from the Department of Atomic Energy, Govt. of India for the measurements at Elettra and the DST-DESY project from Department of Science and Technology, Govt. of India to perform the experiments at PETRA III beamline are thankfully acknowledged. KM acknowledges financial assistance from the Department of Science and Technology, government of India under the J.C. Bose Fellowship program and the Department of Atomic Energy under the DAE-SRC-OI Award program.

\end{document}